\pdfoutput=1
\documentclass[11pt]{article}
\usepackage{amsmath}
\usepackage{graphicx,epstopdf}
\usepackage{cite}
\usepackage{graphicx}
\usepackage[font={small}]{caption}
\usepackage{xcolor}
\usepackage[colorlinks=true,
            linkcolor=red,
            urlcolor=green,
            citecolor=blue]{hyperref}
\definecolor{RawSienna}{cmyk}{0,0.72,1,0.45}
\definecolor{dgreen}{rgb}{0.0,0.42,0.13}
\definecolor{darkblue}{rgb}{0.0, 0.0, 0.55}
\definecolor{cornellred}{rgb}{0.7, 0.11, 0.11}
\definecolor{calpolypomonagreen}{rgb}{0.08, 0.5, 0.5}

\newcommand{\B}{\color{darkblue}}

\usepackage{pbox}
\usepackage{tikz}

\catcode`@=12
\def\beq{\begin{equation}}
\def\eeq{\end{equation}}
\def\bea{\begin{eqnarray}}
\def\eea{\end{eqnarray}}
\makeatletter
\@addtoreset{equation}{section}

\begin{document}
\title{\LARGE \bf Maximal Zero  Textures in Linear and Inverse Seesaw}
\author{{\bf Roopam Sinha\footnote{roopam.sinha@saha.ac.in}, Rome Samanta\footnote{rome.samanta@saha.ac.in}, Ambar Ghosal\footnote{ambar.ghosal@saha.ac.in},  }\\
  Saha Institute of Nuclear Physics, 1/AF Bidhannagar,
  Kolkata 700064, India \\
 }
\maketitle
\begin{abstract}
We investigate Linear and Inverse seesaw mechanisms with maximal zero textures of the constituent matrices subjected to the assumption of non-zero eigenvalues for the neutrino mass matrix $m_\nu$ and charged lepton mass matrix $m_e$. If we restrict to the minimally parametrized non-singular `$m_e$' (i.e., with maximum number of zeros) it gives rise to only 6 possible textures of $m_e$. Non-zero determinant of $m_\nu$ dictates six possible textures of the constituent matrices. We ask in this minimalistic approach, what are the phenomenologically allowed maximum zero textures are possible. It turns out that Inverse seesaw leads to 7 allowed two-zero textures while the Linear seesaw leads to only one. In Inverse seesaw, we show that 2 is the maximum number of independent zeros that can be inserted into $\mu_S$ to obtain all 7 viable two-zero textures of $m_\nu$. On the other hand, in Linear seesaw mechanism, the minimal scheme allows maximum 5 zeros to be accommodated in `$m$' so as to obtain viable effective neutrino mass matrices ($m_\nu$). Interestingly, we find that our minimalistic approach in Inverse seesaw leads to a realization of all the phenomenologically allowed two-zero textures whereas in Linear seesaw only one such texture is viable. Next our numerical analysis shows that none of the two-zero textures give rise to enough CP violation or significant $\delta_{CP}$. Therefore, if $\delta_{CP}=\pi/2$ is established, our minimalistic scheme may still be viable provided we allow more number of parameters in `$m_e$'.
\end{abstract}
\newpage
\section{Introduction}
In Type-I seesaw mechanism the lightness of the observed neutrinos are 
attributed to a seesaw scale around the GUT scale incorporated in the theory. 
In this mechanism, right-handed neutrinos $(\nu_R)$ incorporated in the 
seesaw scale are usually identified with the mass of the 
$\nu_R$: $(M_{\nu_R})$ lightest of which is constrained from leptogenesis as $M_{\text{lightest}}\geq 10^8$ GeV\cite{Davidson:2002qv,Blanchet:2010kw}. Probing the new physics at such a high scale is far beyond the reach of ongoing collider experiments. Moreover, apart from experimental accessibility, a theoretical analysis based on naturalness for a hierarchical $\nu_R$ masses $(M_{R_3}>M_{R_2}>M_{R_1})$ put constraints on them as\cite{Clarke:2015gwa}: 
\begin{equation}
M_{R_1}\leq 4\times 10^7\text{GeV},\hspace{0.2cm}M_{R_2}\leq 7\times 10^7\text{GeV},\hspace{0.2cm} M_{R_3}\leq 3\times 10^7\text{GeV}(\frac{0.05eV}{m_{min}})^{\frac{1}{3}}
\end{equation} where $m_{min}$ is the mass of the lightest neutrino. 
On the other hand, a seesaw scale in the TeV range can be realized in some 
other variants, such as Inverse seesaw, Linear seesaw etc. by paying the 
price in terms of addition of extra singlet neutral fermions into these 
mechanisms which can explain the smallness of neutrino mass by a small 
lepton-number breaking mass matrix. 
The ingredients of these two models incorporate, in addition to the Standard 
Model singlet right-handed neutrinos $\{\nu_{\alpha R}\}$, a set of singlet 
fermions $\{S_{\beta R}\}$, where $\alpha,\beta$ (=1,2,3) are the flavour 
indices. The Yukawa sector of such low energy seesaw mechanism is described by the Lagrangian\cite{Mohapatra:1986bd,Bernabeu:1987gr,Mohapatra:1986aw,Schechter:1981cv,Schechter:1980gr
,Fraser:2014yha,Ghosal:2015lwa,Adhikary:2013mfa,Law:2013gma,Dev:2009aw,
Hernandez:2015tna,Hernandez:2013hea,Hernandez:2014zsa}
\bea
 -\mathcal{L}_{mass}=\overline{\nu_{\alpha L}}m_D^{\alpha\beta}\nu_{\beta R}+M^{\alpha\beta}_R\overline{(\nu_{\alpha R})^c}\nu_{\beta R}+M^{\alpha\beta}_L\overline{\nu_{\alpha L}}(\nu_{\beta L})^c+\mu_S^{\alpha\beta}\overline{(S_{\alpha R})^c}S_{\beta R}\nonumber \\+\overline{\nu_{\alpha L}}M^{\alpha\beta}S_{\beta R}+m^{\alpha\beta}\overline{(\nu_{\alpha R})^c}S_{\beta R}+h.c.
 \eea
\bea 
=\begin{pmatrix}\overline{\nu_{\alpha L}} & \overline{(\nu_{\beta R})^c} & \overline{(S_{\beta R})^c}\end{pmatrix}\begin{pmatrix}M_L & m_D & M\\ m_D^T & M_R & m\\ M^T & m & \mu_S\end{pmatrix}\begin{pmatrix}(\nu_{\beta L})^c\\ \nu_{\beta R}\\ S_{\beta R}\end{pmatrix}+h.c.
\eea
 where $m_D$, $M$, $m$ (since it is due to combination of two different 
fields)  are the Dirac type and the rest are the Majorana type mass matrices. 
Usually the Linear seesaw mechanism is facilitated with the exclusion of all 
other lepton number violating mass terms expect `$m$' whereas in Inverse 
seesaw mechanism both $\mu_S$ and $m$ contain lepton number violating mass terms. Thus for Linear seesaw, we consider diagonal entries $M_L=M_R=\mu_S=0$ and for Inverse seesaw, $M_L=M_R=M=0$.
 Therefore, the low energy effective neutrino mass matrix in Linear seesaw\cite{Barr:2003nn,Hettmansperger:2011bt,Hirsch:2009mx,Chakraborty:2014hfa,Wang:2015saa} can be written as 
\bea 
 m_\nu &\approx & -M(m^{-1}m_D^T)-[M(m^{-1}m_D^T)]^T
 \eea 
and accordingly in Inverse seesaw it turns out as 
\bea
m_\nu &\approx & m_D m^{-1}\mu_S ( m_D m^{-1})^T. \label{inv}
\eea
Now as there are fewer number of experimental constraints, a fruitful 
approach is to minimize the number of parameters in the Lagrangian. 
Popular paradigm is to consider some symmetry in the Lagrangian that 
reduces the number of parameters or to assume texture zeros 
(which are also dictated by some underlying symmetry) in the fundamental 
mass matrices.
\vskip 0.1in
\noindent
In our present work we investigate both the low energy seesaw  mechanisms mentioned earlier, incorporating the idea of maximal zero textures\cite{Frampton:2002yf,
Whisnant:2014yca,Ludl:2014axa,Grimus:2013qea,Liao:2013saa,Fritzsch:2011qv,Merle:2006du,
Wang:2014dka,Lavoura:2004tu,Kageyama:2002zw,Wang:2014vua,Baek:2015mna,4zero1,
Choubey:2008tb,
4zero2,4zero3,4zero4,Samanta:2015hxa,Samanta:2015oqa,Adhikary:2012kb} 
subjected to the criterion of non-zero eigenvalues of the charged lepton 
($m_e$) and effective neutrino mass matrix ($m_\nu$). We investigate the 
viable textures of $m_\nu$ with maximum number of zeros that can be 
accommodated with the current data. Our methodology is as follows:\\
i) First we explore to find out a minimal texture of charged lepton mass matrix ($m_e$) which gives rise to three distinct nonzero eigenvalues, i.e, minimum number of parameters necessary to obtain det($m_em_e^{\dagger}$)$\neq$ 0. The textures obtained are such that they do not contribute to $U_{PMNS}$.\\
ii) Next we assume all the three light neutrino eigenvalues of $m_\nu$ are non-zero i.e., det$(m_\nu)\neq 0$. The Linear seesaw formula implies that $m_D$, $m$ and $M$ are also non-singular. This fact unambiguously determines the possible minimal textures of $m_D$, $m$ and $M$. In the Inverse seesaw, the same criterion fixes the minimal textures of $m_D$, $\mu_S$ and $m$.\\
iii) Fixing a particular minimal structure of $m_D$ and $M$ in Linear seesaw (or $m_D$ and $m$ in Inverse Seesaw), we systematically explore to obtain the minimal texture of the matrix $m$ (in Linear Seesaw) and $\mu_S$ in Inverse Seesaw by putting zeros in different entries, for the case of Linear (Inverse) seesaw.\\
iv) Following, we utilize the Frampton and Glashow and Marfatia condition\cite{Frampton:2002yf} to eliminate emerged unphysical effective neutrino matrices ($m_\nu$).\\
v) Finally, we explore numerically the parameter space of the survived matrices utilizing the neutrino oscillation global fit data and predict $\Sigma_im_i$, $|m_{11}|$, $J_{CP}$, $\delta_{CP}$ along with the hierarchical structure of neutrino masses.\\
The paper is organized as follows: Sec. \ref{s2} contains minimally parametrized charged lepton mass matrices $m_e$ and it is obtained that they do not contribute to $U_{PMNS}$. Effective neutrino mass matrices arising from texture zeros in Linear seesaw is discussed in Sec. \ref{s3}. The same analysis for Inverse seesaw in presented in Sec. \ref{s4}. Sec. \ref{s5} contains the summary of the present work.
\section{The minimal charged lepton basis}\label{s2}
In general, the charged lepton mass matrix has the form \begin{equation}m_e=\begin{pmatrix}A^\prime e^{ia^\prime} & B^\prime e^{ib^\prime}  & C^\prime e^{ic^\prime}\\ D^\prime e^{id^\prime} & E^\prime e^{ie^\prime} & F^\prime e^{if^\prime}\\ G^\prime e^{ig^\prime } & H^\prime e^{ih^\prime} & K^\prime e^{ik^\prime}\end{pmatrix}\label{clep}.\end{equation} We look for maximum zero textures (minimum number of parameters) of $m_e$ such that $\det(m_em_e^\dag)\neq 0$ (or non-zero eigenvalues for $m_e$). A careful inspection of the determinant $det(m_em_e^\dag)$ reveals six stringent possibilities and are presented accordingly in Table \ref{t1}.
\begin{table}[!h]
\caption{Minimal textures of the charged lepton mass matrix $m_e$} \label{t1}
\begin{center}
\begin{tabular}{|p{5.5cm}|p{5.5cm}|p{5.5cm}|}
\hline
$m_e^{(1)}=\begin{pmatrix}A^\prime e^{ia^\prime} & 0 & 0\\0 & B^\prime e^{ib^\prime} & 0\\ 0 & 0 & C^\prime e^{ic^\prime}\end{pmatrix}$ & 
$m_e^{(2)}=\begin{pmatrix}0 & 0 & A^\prime e^{ia^\prime}\\0 & B^\prime e^{ib^\prime} & 0\\C^\prime e^{ic^\prime} & 0 & 0\end{pmatrix}$ &
$m_e^{(3)}=\begin{pmatrix}A^\prime e^{ia^\prime} & 0 & 0\\0 & 0 & B^\prime e^{ib^\prime}\\ 0 & C^\prime e^{ic^\prime} & 0\end{pmatrix}$\\
 \hline
 $m_e^{(4)}=\begin{pmatrix}0 & 0 & A^\prime e^{ia^\prime}\\B^\prime e^{ib^\prime} & 0 & 0\\0 & C^\prime e^{ic^\prime} & 0\end{pmatrix}$ & 
$m_e^{(5)}=\begin{pmatrix}0 & A^\prime e^{ia^\prime} & 0\\0 & 0 & B^\prime e^{ib^\prime}\\C^\prime e^{ic^\prime} & 0 & 0\end{pmatrix}$ &
$m_e^{(6)}=\begin{pmatrix}0 & A^\prime e^{ia^\prime} & 0\\B^\prime e^{ib^\prime} & 0 & 0\\0 & 0 & C^\prime e^{ic^\prime}\end{pmatrix}$\\
 \hline
\end{tabular}
\end{center}
\end{table} 
Interestingly, for all these matrices, $m_em_e^\dag$ is diagonal. The matrix ($U_l$) that diagonalizes $m_em_e^\dag$ is an unit matrix and therefore, the mixing arises only from the neutrino sector of the Lagrangian since $U_{PMNS}={U_l}^{\dagger}U_{\nu}$.

\section{Texture zeros in Linear seesaw}\label{s3}

If A is an invertible $n\times n$ square matrix, and B and C are $n\times m$ matrices, then \begin{equation}
det(A+BC^T)=det(I_m+C^TA^{-1}B)detA
\end{equation} Therefore, if we assume det$(m_\nu)\neq 0$, the Linear Seesaw formula implies that $det(Mm^{-1}m_D^T)\neq 0$. Therefore, $M$, $m^{-1}$ and 
$m_D^T$ must be non-singular. Since, for a matrix A, $det(A)=det(A^T)$, 
$det(A^{-1})=1/det(A)$, we obtain that $m_D$, $M$ and $m$ must be non-singular. These leads to the following textures of $m_D$, $M$ and $m$ and are shown in 
Table 2, Table 3 and Table 4 respectively.
 \begin{table}[!h]
\caption{Minimal (6-zero) textures of $m_D$ with $det(m_D)\neq 0$} \label{t2}
\begin{center}
\begin{tabular}{|p{5.5cm}|p{5.5cm}|p{5.5cm}|}
\hline
\multicolumn{3}{|c|}{{\bf Minimal (6-zero) textures of $m_D$ with $det(m_D)\neq 0$}}\\
\hline
$m_D^{(1)}=\begin{pmatrix}Ae^{ia} & 0 & 0\\0 & Be^{ib} & 0\\ 0 & 0 & Ce^{ic}\end{pmatrix}$ & 
$m_D^{(2)}=\begin{pmatrix}0 & 0 & Ae^{ia}\\0 & Be^{ib} & 0\\Ce^{ic} & 0 & 0\end{pmatrix}$ &
$m_D^{(3)}=\begin{pmatrix}Ae^{ia} & 0 & 0\\0 & 0 & Be^{ib}\\ 0 & Ce^{ic} & 0\end{pmatrix}$\\
 \hline
 $ m_D^{(4)}=\begin{pmatrix}0 & 0 & Ae^{ia}\\Be^{ib} & 0 & 0\\0 & Ce^{ic} & 0\end{pmatrix}$ & 
$m_D^{(5)}=\begin{pmatrix}0 & Ae^{ia} & 0\\0 & 0 & Be^{ib}\\Ce^{ic} & 0 & 0\end{pmatrix}$ &
$m_D^{(6)}=\begin{pmatrix}0 & Ae^{ia} & 0\\Be^{ib} & 0 & 0\\0 & 0 & Ce^{ic}\end{pmatrix}$\\
 \hline
\end{tabular}
\end{center}
\end{table}

 \begin{table}[!h]
\caption{Minimal (6-zero) textures of $M$ with $det(M)\neq 0$} \label{t29}
\begin{center}
\begin{tabular}{|p{5.5cm}|p{5.5cm}|p{5.5cm}|}
\hline
\multicolumn{3}{|c|}{{\bf Minimal (6-zero) textures of $M$ with $det(M)\neq 0$}}\\
\hline
$M^{(1)}=\begin{pmatrix}Xe^{ix} & 0 & 0\\0 & Ye^{iy} & 0\\ 0 & 0 & Ze^{iz}\end{pmatrix}$ & 
$M^{(2)}=\begin{pmatrix}0 & 0 & Xe^{ix}\\0 & Ye^{iy} & 0\\Ze^{iz} & 0 & 0\end{pmatrix}$ &
$M^{(3)}=\begin{pmatrix}Xe^{ix} & 0 & 0\\0 & 0 & Ye^{iy}\\ 0 & Ze^{iz} & 0\end{pmatrix}$\\
 \hline
 $M^{(4)}=\begin{pmatrix}0 & 0 & Xe^{ix}\\Ye^{iy} & 0 & 0\\0 & Ze^{iz} & 0\end{pmatrix}$ & 
$M^{(5)}=\begin{pmatrix}0 & Xe^{ix} & 0\\0 & 0 & Ye^{iy}\\Ze^{iz} & 0 & 0\end{pmatrix}$ &
$M^{(6)}=\begin{pmatrix}0 & Xe^{ix} & 0\\Ye^{iy} & 0 & 0\\0 & 0 & Ze^{iz}\end{pmatrix}$\\
 \hline
\end{tabular}
\end{center}
\end{table}
 \begin{table}[!h]
\caption{Minimal (6-zero) textures of $m$ with $det(m)\neq 0$} \label{t30}
\begin{center}
\begin{tabular}{|p{5.5cm}|p{5.5cm}|p{5.5cm}|}
\hline
\multicolumn{3}{|c|}{{\bf Minimal (6-zero) textures of $m$ with $det(m)\neq 0$}}\\
\hline
$m^{(1)}=\begin{pmatrix}Pe^{ip} & 0 & 0\\0 & Qe^{iq} & 0\\ 0 & 0 & Re^{ir}\end{pmatrix}$ & 
$m^{(2)}=\begin{pmatrix}0 & 0 & Pe^{ip}\\0 & Qe^{iq} & 0\\Re^{ir} & 0 & 0\end{pmatrix}$ &
$m^{(3)}=\begin{pmatrix}Pe^{ip} & 0 & 0\\0 & 0 & Qe^{iq}\\ 0 & Re^{ir} & 0\end{pmatrix}$\\
 \hline
 $m^{(4)}=\begin{pmatrix}0 & 0 & Pe^{ip}\\Qe^{iq} & 0 & 0\\0 & Re^{ir} & 0\end{pmatrix}$ & 
$m^{(5)}=\begin{pmatrix}0 & Pe^{ip} & 0\\0 & 0 & Qe^{iq}\\Re^{ir} & 0 & 0\end{pmatrix}$ &
$m^{(6)}=\begin{pmatrix}0 & Pe^{ip} & 0\\Qe^{iq} & 0 & 0\\0 & 0 & Re^{ir}\end{pmatrix}$\\
 \hline
\end{tabular}
\end{center}
\end{table}
\newpage
\subsection{ Effective $m_{\nu}$ in Linear seesaw}
 Our basic requirement is to admit a 
structure of $m_\nu$ is based on the result of 
Ref.\cite{Frampton:2002yf} in which it is shown that to obtain a 
phenomenologically viable $m_\nu$, the number of independent zeros should be 
atleast two. Keeping such criterion in view we start with maximum number of zeros in the matrix $m$ for a given $m_D$ and $M$. It turns out that for 6 zeros in $m$, all the emerged $m_\nu$ has either three or more independent zeros. Discarding such textures, therefore, we start with five zero textures of $m$. In general, there are 126 possible 5-zero textures of $m$. But implementation of Linear 
seesaw, as well as our demand of non-zero eigenvalues of $m_\nu$ requires $m$ to be non-singular. The requirement drastically reduces the number of non-singular 5-zero textures of $m$ to 36 which are  presented in Table \ref{5zm}.

\begin{table}[!h]
\caption{Viable 5-zero textures of $m$} \label{5zm}
\begin{center}
\begin{tabular}{|p{5.5cm}|p{5.5cm}|p{5.5cm}|}
\hline
\multicolumn{3}{|c|}{{\bf 5 zero textures of $m$}}\\
\hline
$m^1=\begin{pmatrix}Pe^{ip} & Se^{is} & 0\\0 & Qe^{iq} & 0\\0 & 0 & Re^{ir}\end{pmatrix}$ & 
$m^2=\begin{pmatrix}Pe^{ip} & 0 & Se^{is}\\0 & Qe^{iq} & 0\\0 & 0 & Re^{ir}\end{pmatrix}$ &
$m^3=\begin{pmatrix}Pe^{ip} & 0 & 0\\Se^{is} & Qe^{iq} & 0\\0 & 0 & Re^{ir}\end{pmatrix}$\\
\hline
$m^4=\begin{pmatrix}Pe^{ip} & 0 & 0\\0 & Qe^{iq} & Se^{is}\\0 & 0 & Re^{ir}\end{pmatrix}$ & 
$m^5=\begin{pmatrix}Pe^{ip} & 0 & 0\\0 & Qe^{iq} & 0\\Se^{is} & 0 & Re^{ir}\end{pmatrix}$ &
$m^6=\begin{pmatrix}Pe^{ip} & 0 & 0\\0 & Qe^{iq} & 0\\0 & Se^{is} & Re^{ir}\end{pmatrix}$\\
\hline
 $m^7=\begin{pmatrix}Se^{is} & Pe^{ip} & 0\\Qe^{iq} & 0 & 0\\0 & 0 & Re^{ir}\end{pmatrix}$ & 
$m^8=\begin{pmatrix}0 & Pe^{ip} & Se^{is}\\Qe^{iq} & 0 & 0\\0 & 0 & Re^{ir}\end{pmatrix}$ &
$m^9=\begin{pmatrix}0 & Pe^{ip} & 0\\Qe^{iq} & Se^{is} & 0\\0 & 0 & Re^{ir}\end{pmatrix}$\\
 \hline
 $m^{10}=\begin{pmatrix}0 & Pe^{ip} & 0\\Qe^{iq} & 0 & Se^{is}\\0 & 0 & Re^{ir}\end{pmatrix}$ & 
$m^{11}=\begin{pmatrix}0 & Pe^{ip} & 0\\Qe^{iq} & 0 & 0\\Se^{is} & 0 & Re^{ir}\end{pmatrix}$ &
$m^{12}=\begin{pmatrix}0 & Pe^{ip} & 0\\Qe^{iq} & 0 & 0\\0 & Se^{is} & Re^{ir}\end{pmatrix}$\\
 \hline
 $m^{13}=\begin{pmatrix}Se^{is} & 0 & Pe^{ip} \\Qe^{iq} & 0 & 0\\0 & Re^{ir} & 0\end{pmatrix}$ & 
$m^{14}=\begin{pmatrix}0 & Se^{is} & Pe^{ip} \\Qe^{iq} & 0 & 0\\0 & Re^{ir} & 0\end{pmatrix}$ &
$m^{15}=\begin{pmatrix}0 & 0 & Pe^{ip} \\Qe^{iq} & Se^{is} & 0\\0 & Re^{ir} & 0\end{pmatrix}$\\
 \hline
 $m^{16}=\begin{pmatrix}0 & 0 & Pe^{ip} \\Qe^{iq} & 0 & Se^{is}\\0 & Re^{ir} & 0\end{pmatrix}$ & 
$m^{17}=\begin{pmatrix}0 & 0 & Pe^{ip} \\Qe^{iq} & 0 & 0\\Se^{is} & Re^{ir} & 0\end{pmatrix}$ &
$m^{18}=\begin{pmatrix}0 & 0 & Pe^{ip} \\Qe^{iq} & 0 & 0\\0 & Re^{ir} & Se^{is}\end{pmatrix}$\\
 \hline
 $m^{19}=\begin{pmatrix}Se^{is} & Pe^{ip} & 0 \\0 & 0 & Qe^{iq}\\Re^{ir} & 0 & 0\end{pmatrix}$ & 
$m^{20}=\begin{pmatrix}0 & Pe^{ip} & Se^{is} \\0 & 0 & Qe^{iq}\\Re^{ir} & 0 & 0\end{pmatrix}$ &
$m^{21}=\begin{pmatrix}0 & Pe^{ip} & 0 \\Se^{is} & 0 & Qe^{iq}\\Re^{ir} & 0 & 0\end{pmatrix}$\\
 \hline
 $m^{22}=\begin{pmatrix}0 & Pe^{ip} & 0 \\0 & Se^{is} & Qe^{iq}\\Re^{ir} & 0 & 0\end{pmatrix}$ & 
$m^{23}=\begin{pmatrix}0 & Pe^{ip} & 0 \\0 & 0 & Qe^{iq}\\Re^{ir} & Se^{is} & 0\end{pmatrix}$ &
$m^{24}=\begin{pmatrix}0 & Pe^{ip} & 0 \\0 & 0 & Qe^{iq}\\Re^{ir} & 0 & Se^{is}\end{pmatrix}$\\
 \hline
 $m^{25}=\begin{pmatrix}Pe^{ip} & Se^{is} & 0 \\0 & 0 & Qe^{iq}\\0 & Re^{ir} & 0\end{pmatrix}$ & 
$m^{26}=\begin{pmatrix}Pe^{ip} & 0 & Se^{is}\\0 & 0 & Qe^{iq}\\0 & Re^{ir} & 0\end{pmatrix}$ &
$m^{27}=\begin{pmatrix}Pe^{ip} & 0 & 0 \\Se^{is} & 0 & Qe^{iq}\\0 & Re^{ir} & 0\end{pmatrix}$\\
 \hline
 $m^{28}=\begin{pmatrix}Pe^{ip} & 0 & 0 \\0 & Se^{is} & Qe^{iq}\\0 & Re^{ir} & 0\end{pmatrix}$ & 
$m^{29}=\begin{pmatrix}Pe^{ip} & 0 & 0 \\0 & 0 & Qe^{iq}\\Se^{is} & Re^{ir} & 0\end{pmatrix}$ &
$m^{30}=\begin{pmatrix}Pe^{ip} & 0 & 0 \\0 & 0 & Qe^{iq}\\0 & Re^{ir} & Se^{is}\end{pmatrix}$\\
 \hline
 $m^{31}=\begin{pmatrix}Se^{is} & 0 & Pe^{ip} \\0 & Qe^{iq} & 0\\Re^{ir} & 0 & 0\end{pmatrix}$ & 
$m^{32}=\begin{pmatrix}0 & Se^{is} & Pe^{ip} \\0 & Qe^{iq} & 0\\Re^{ir} & 0 & 0\end{pmatrix}$ &
$m^{33}=\begin{pmatrix}0 & 0 & Pe^{ip} \\Se^{is} & Qe^{iq} & 0\\Re^{ir} & 0 & 0\end{pmatrix}$\\
 \hline
 $m^{34}=\begin{pmatrix}0 & 0 & Pe^{ip} \\0 & Qe^{iq} & Se^{is}\\Re^{ir} & 0 & 0\end{pmatrix}$ & 
$m^{35}=\begin{pmatrix}0 & 0 & Pe^{ip} \\0 & Qe^{iq} & 0\\Re^{ir} & Se^{is} & 0\end{pmatrix}$ &
$m^{36}=\begin{pmatrix}0 & 0 & Pe^{ip} \\0 & Qe^{iq} & 0\\Re^{ir} & 0 & Se^{is}\end{pmatrix}$\\
 \hline
\end{tabular}
\end{center}
\end{table} 
\clearpage

Interestingly, only the combinations given in Table (Table \ref{t5}) gives rise to phenomenologically viable $m_\nu$ and all of them 
leads to a single generic structure as \begin{equation}
\begin{pmatrix}
\times & \times & \times\\
\times & 0 & \times\\
\times & \times & 0
\end{pmatrix}.\label{eqq}
\end{equation} All the other combinations are discarded because they either 
lead to 3 independent zeros or 2 independent zeros that are not consistent with the current data. For a compact view we present a Table (Table \ref{t5}) that contains all the allowed combinations of $m_D$, $M$ and $m$.\\

\begin{table}[!h]
\caption{Compositions for Allowed Realizations of $m_\nu$ }\label{t5}
\begin{center}
\begin{tabular}{|p{1cm}|p{1.5cm}|p{1.5cm}|p{1.5cm}|p{1.5cm}|p{1.5cm}|p{1.5cm}|}
\cline{2-7}
\multicolumn{1}{c|}{}  & \multicolumn{6}{c|}{$M$ $\rightarrow$} \\ 
\hline
$m_D$ $\downarrow$ & $M^{(1)}$ &$M^{(2)}$ & $M^{(3)}$ & $M^{(4)}$ & $M^{(5)}$ & $M^{(6)}$\\
\hline
$ m_D^{(1)} $ & $m^{16},m^{23}$ & $m^{12},m^{17}$ &$m^{9},m^{36}$ & $m^{5},m^{22}$ & $m^{3},m^{18}$ & $m^{29},m^{34}$\\
\hline
$m_D^{(2)}$&$m^{10},m^{25}$&$m^{14},m^{21}$&$m^{2},m^{15}$&$m^{28},m^{31}$& $m^{8},m^{33}$ & $m^{4},m^{19}$\\
\hline
$ m_D^{(3)} $ & $m^{9},m^{36}$ & $m^{5},m^{22}$ &$m^{16},m^{23}$ & $m^{12},m^{27}$& $m^{29},m^{34}$ & $m^{3},m^{18}$\\
\hline
$m_D^{(4)}$&$m^{2},m^{15}$&$m^{28},m^{31}$&$m^{10},m^{25}$&$m^{14},m^{21}$& $m^{4},m^{19}$ & $m^{8},m^{33}$\\
\hline
$ m_D^{(5)} $ & $m^{1},m^{24}$ & $m^{11},m^{32}$ &$m^{26},m^{35}$ & $m^{6},m^{13}$& $m^{17},m^{20}$ & $m^{7},m^{30}$\\
\hline
$m_D^{(6)}$&$m^{26},m^{35}$&$m^{6},m^{13}$&$m^{1},m^{24}$&$m^{11},m^{31}$& $m^{7},m^{30}$ & $m^{17},m^{20}$\\
\hline
\end{tabular}
\end{center}
\end{table}  
Moreover, the transpose and inverse of the above 3 zero textures of $m$ are such that they yield same 3 zero textures. Thus the seesaw formula implies if 5-zero textures are assumed in $M$ (instead in $m$) and 6 zeros in $m$ and $m_D$, the resulting textures will be identical to the case where $m_D$ is assumed to have 5-zeros and remaining matrices $m, M$ contain 6 zeros. It turns out that none of the permutations generate two-zero textures other than the one already obtained above.
\subsubsection{Parametrization and phase rotation}

To be explicit, we parametrize one set of combination 
($m^3_D$ and $m^9$) and ($m^3_D$ and $m^{36}$) which gives rise to $m_\nu$ 
given in Eq.\eqref{eqq}. To extract the relevant phases out of these allowed $m_\nu$, let us parametrize them in a generic way as
\bea
 m_\nu &=&\begin{pmatrix} K_1e^{ik_1} & K_2e^{ik_2} & K_3e^{ik_3}\\K_2e^{ik_2} & 0 & K_4e^{ik_4} \\K_3e^{ik_3} & K_4e^{ik_4} & 0\end{pmatrix}
\eea
where for $m_D^3$ with $m^9$ combination:
\bea
K_1e^{ik_1}=\frac{2 A e^{i a-ip-i q+i s+i x} S X}{PQ},\hspace{0.5cm}K_2e^{ik_2}= -\frac{A e^{i a-i p+i y} Y}{P},\nonumber \\
K_3e^{ik_3}=-\frac{C e^{i c-i q+i x} X}{Q},\hspace{0.5cm} K_4e^{ik_4}=-\frac{B e^{i b-i r+i z} Z}{R}.
\eea
and for $m_D^3$ with $m^{36}$ combination:
\bea
K_1e^{ik_1}= \frac{2 A e^{i a-ip-i r+i s+i x} S X}{PR},\hspace{0.5cm}K_2e^{ik_2}=-\frac{B e^{i b-i r+i x} X}{R},\nonumber \\K_3e^{ik_3}=-\frac{A e^{i a-i p+i z} Z}{P},\hspace{0.5cm}  K_4e^{ik_4}=-\frac{C e^{i c-i q+i y} Y}{Q}.\label{a1}
\eea

\section{ Texture zeros in Inverse seesaw}\label{s4}

As before we consider the minimal non-singular textures of $m_D$ (presented 
in Table \ref{t2}) and  $m$ (Table \ref{t81})  whereas the minimal texture of $\mu_S$ contains only two-independent complex parameters due to its anti-symmetry and given by the 3 possible textures presented in Table \ref{t50}.

\begin{table}[!h]
\caption{Minimal ($6$-zero) textures of $m$} \label{t81}
\begin{center}
\begin{tabular}{|p{5.5cm}|p{5.5cm}|p{5.5cm}|}
\hline
\multicolumn{3}{|c|}{{\bf Minimal ($6$-zero) textures of $m$}}\\
\hline
$m^{(1)}=\begin{pmatrix}Xe^{ix} & 0 & 0\\0 & Ye^{iy} & 0\\ 0 & 0 & Ze^{iz}\end{pmatrix}$ & 
$m^{(2)}=\begin{pmatrix}0 & 0 & Xe^{ix}\\0 & Ye^{iy} & 0\\Ze^{iz} & 0 & 0\end{pmatrix}$ &
$m^{(3)}=\begin{pmatrix}Xe^{ix} & 0 & 0\\0 & 0 & Ye^{iy}\\ 0 & Ze^{iz} & 0\end{pmatrix}$\\
 \hline
 $ m^{(4)}=\begin{pmatrix}0 & 0 & Xe^{ix}\\Ye^{iy} & 0 & 0\\0 & Ze^{iz} & 0\end{pmatrix}$ & 
$m^{(5)}=\begin{pmatrix}0 & Xe^{ix} & 0\\0 & 0 & Ye^{iy}\\Ze^{iz} & 0 & 0\end{pmatrix}$ &
$m^{(6)}=\begin{pmatrix}0 & Xe^{ix} & 0\\Ye^{iy} & 0 & 0\\0 & 0 & Ze^{iz}\end{pmatrix}$\\
 \hline
\end{tabular}
\end{center}
\end{table}
\begin{table}[!h]
\caption{Minimal ($4$-independent zero) textures of $\mu_S$} \label{t50}
\begin{center}
\begin{tabular}{|p{5.5cm}|p{5.5cm}|p{5.5cm}|}
\hline
\multicolumn{3}{|c|}{{\bf Minimal ($4$-independent zero) textures of $\mu_S$}}\\
\hline
$\mu_S^{(1)}=\begin{pmatrix}0 & 0 & Re^{ir}\\0 & Se^{is} & 0\\Re^{ir} & 0 & 0\end{pmatrix}$ & 
$\mu_S^{(2)}=\begin{pmatrix}Pe^{ip} & 0 & 0\\0 & 0 & Te^{it}\\0 & Te^{it} & 0\end{pmatrix}$ &
$\mu_S^{(3)}=\begin{pmatrix}0 & Qe^{iq} & 0\\Qe^{iq} & 0 & 0\\ 0 & 0 & Ve^{iv}\end{pmatrix}$\\
 \hline
\end{tabular}
\end{center}
\end{table}

However, it turns out that if the number of zeros in $\mu_S$ is greater than 3 (as in Table \ref{t50}), all the emerged $m_\nu$ contain 3 or more independent zeros and hence discarded. Therefore, to obtain viable structures of $m_\nu$ we stick with the non-singular 2 zero textures of $\mu_S$ and are presented in Table \ref{t6}. Interestingly, unlike Linear seesaw, we note that Inverse seesaw leads to all the 7 viable two-zero textures ($m_\nu^1$ - $m_\nu^7$) of $m_\nu$ given in Ref.\cite{Frampton:2002yf}. In a compact way, in Table \ref{t7} we present all the combinations that generate these textures of $m_\nu$.\\
\begin{table}[h!]
\caption{2-independent zero textures for $\mu_S$ } \label{t6}
\begin{center}
\begin{tabular}{|p{5cm}|p{5cm}|p{5cm}|}
\hline
\multicolumn{3}{|c|}{{\bf 2-independent-zero textures for $\mu_S$}}\\
\hline
$\mu_S^1=\begin{pmatrix}0 & 0  & Re^{ir}\\0 & Se^{is} & Te^{it}\\Re^{ir} & Te^{it} & Ve^{iv}\end{pmatrix}$ &
$\mu_S^2=\begin{pmatrix}0 & Qe^{iq}  & Re^{ir}\\Qe^{iq} & Se^{is} & 0\\Re^{ir} & 0 & Ve^{iv}\end{pmatrix}$ &
$\mu_S^3=\begin{pmatrix}0 & Qe^{iq}  & 0\\Qe^{iq} & Se^{is} & Te^{it}\\0 & Te^{it} & Ve^{iv}\end{pmatrix}$\\
\hline
$\mu_S^4=\begin{pmatrix}Pe^{ip} & 0  & Re^{ir}\\0 & 0 & Te^{it}\\Re^{ir} & Te^{it} & Ve^{iv}\end{pmatrix}$ &
$\mu_S^5=\begin{pmatrix}Pe^{ip} & Qe^{iq}  & Re^{ir}\\Qe^{iq} & 0 & 0\\Re^{ir} & 0 & Ve^{iv}\end{pmatrix}$ &
$\mu_S^6=\begin{pmatrix}Pe^{ip} & Qe^{iq}  & 0\\Qe^{iq} & 0 & Te^{it}\\0 & Te^{it} & Ve^{iv}\end{pmatrix}$\\
\hline
$\mu_S^7=\begin{pmatrix}Pe^{ip} & 0  & Re^{ir}\\0 & Se^{is} & Te^{it}\\Re^{ir} & Te^{it} & 0\end{pmatrix}$ &
$\mu_S^8=\begin{pmatrix}Pe^{ip} & Qe^{iq}  & Re^{ir}\\Qe^{iq} & Se^{is} & 0\\Re^{ir} & 0 & 0\end{pmatrix}$ &
$\mu_S^9=\begin{pmatrix}Pe^{ip} & Qe^{iq}  & 0\\Qe^{iq} & Se^{is} & Te^{it}\\0 & Te^{it} & 0\end{pmatrix}$\\
\hline
$\mu_S^{10}=\begin{pmatrix}Pe^{ip} & Qe^{iq}  & Re^{ir}\\Qe^{iq} & 0 & Te^{it}\\Re^{ir} & Te^{it} & 0\end{pmatrix}$ &
$\mu_S^{11}=\begin{pmatrix}0 & Qe^{iq}  & Re^{ir}\\Qe^{iq} & Se^{is} & Te^{it}\\Re^{ir} & Te^{it} & 0\end{pmatrix}$ &
$\mu_S^{12}=\begin{pmatrix}0 & Qe^{iq}  & Re^{ir}\\Qe^{iq} & 0 & Te^{it}\\Re^{ir} & Te^{it} & Ve^{iv}\end{pmatrix}$\\
\hline
\end{tabular}
\end{center}
\end{table}
\newpage
\begin{table}[!h]
\caption{Compositions for Realization of two-zero $m_\nu$ textures with $m_D=m_D^{(1)}$}\label{t7}
\begin{center}
\begin{tabular}{|p{0.7cm}|p{0.7cm}|p{0.7cm}|p{0.7cm}|p{0.7cm}|p{0.7cm}|p{0.7cm}|}
\cline{2-7}
\multicolumn{1}{c|}{}  & \multicolumn{6}{c|}{$m$ $\rightarrow$} \\ \hline
$\mu_S$ $\downarrow$ & $m^1$ &$m^2$ & $m^3$ & $m^4$ & $m^5$ & $m^6$\\
\hline
$\mu_S^1 $ & $m_\nu^1$ & $\times$ &$m_\nu^2$ & $m_\nu^6$ & $\times$ &$m_\nu^5$\\
\hline
$\mu_S^2$ & $\times$ & $m_\nu^4$ & $\times$ & $m_\nu^4$ & $m_\nu^3$ & $m_\nu^3$\\
\hline
$\mu_S^3 $ & $m_\nu^2$ & $m_\nu^6$ & $m_\nu^1$ & $\times$ & $\times$ &$\times$\\
\hline
$\mu_S^4 $ & $m_\nu^5$ & $\times$ & $m_\nu^6$ & $m_\nu^2$ & $m_\nu^5$ & $m_\nu^1$\\
\hline
$\mu_S^5 $ & $\times$ & $m_\nu^5$ & $\times$ & $m_\nu^1$ & $\times$ &$m_\nu^2$\\
\hline
$\mu_S^6 $ & $m_\nu^3$ & $m_\nu^3$ & $m_\nu^4$ & $\times$ & $m_\nu^6$ &$\times$\\
\hline
$\mu_S^7 $ & $m_\nu^4$ & $\times$ & $m_\nu^3$ & $m_\nu^3$ & $m_\nu^4$ & $m_\nu^4$\\
\hline
$\mu_S^8 $ & $\times$ & $m_\nu^1$ & $\times$ & $m_\nu^5$ & $m_\nu^2$ &$m_\nu^6$\\
\hline
$\mu_S^9 $ & $m_\nu^6$ & $m_\nu^2$ & $m_\nu^5$ & $\times$ & $m_\nu^1$ & $\times$\\
\hline
$\mu_S^{10} $ & $m_\nu^7$ & $\times$ & $m_\nu^7$ & $\times$ & $\times$ &$\times$\\
\hline
$\mu_S^{11} $ & $\times$ & $\times$ & $\times$ & $m_\nu^7$ & $\times$ &$m_\nu^7$\\
\hline
$\mu_S^{12} $ & $\times$ & $m_\nu^7$ & $\times$ & $\times$ & $m_\nu^7$ & $\times$\\
\hline
\end{tabular}
\end{center}
\end{table} 
Similar to Table \ref{t7}, five more tables can be obtained for $m_D^{(2)}-m_D^{(6)}$. However, all those combinations also lead to all seven possible two-zero textures but with different combinations of $m_D$, $m$ and $\mu_S$. We are not listing all these tables.

\subsection{Effective $m_\nu$ and its parametrization}
\begin{table}[!h]
\caption{Effective allowed $m_\nu$ from Inverse seesaw} \label{t20}
\begin{center}
\begin{tabular}{|p{5.7cm}|p{5.5cm}|p{5.0cm}|}
\hline
\multicolumn{3}{|c|}{{\bf Effective allowed $m_\nu$ from Inverse seesaw}}\\
\hline
$\hspace{2.8cm}m_\nu$ & 
\text{Phase rotated} $m_\nu$ & \text{Parametrization}\\
\hline
$m_\nu^1=\begin{pmatrix}0 & 0 & K_1e^{ik_1}\\0 & K_2e^{ik_2} & K_3e^{ik_3}\\ K_1e^{ik_1} & K_3e^{ik_3} & K_4e^{ik_4}\end{pmatrix}$ &  
$m_\nu^1=m_0\begin{pmatrix}0 & 0 & 1\\ 0 & y_1 & y_2\\0 & y_2 & y_3e^{i\alpha}\end{pmatrix}$ &
 $m_0=K_1,$ $K_2/K_1=y_1,$ $K_3/K_1=y_2,$ $K_4/K_1=y_3,$ $\alpha=(k_2-2k_3+k_4)$\\
 \hline
$m_\nu^2=\begin{pmatrix}0 & K_1e^{ik_1} & 0\\K_1e^{ik_1} & K_2e^{ik_2} & K_3e^{ik_3}\\ 0 & K_3e^{ik_3} & K_4e^{ik_4}\end{pmatrix}$ & 
$m_\nu^2=m_0\begin{pmatrix}0 & 1 & 0\\ 1 & y_1 & y_2\\0 & y_2 & y_3e^{i\alpha}\end{pmatrix}$ &
 $m_0=K_1,$ $K_2/K_1=y_1,$ $K_3/K_1=y_2,$ $K_4/K_1=y_3,$ $\alpha=(k_2-2k_3+k_4)$\\
 \hline
$m_\nu^3=\begin{pmatrix}K_1e^{ik_1} & K_2e^{ik_2} & 0\\K_2e^{ik_2} & 0 & K_3e^{ik_3}\\ 0 & K_3e^{ik_3} & K_4e^{ik_4}\end{pmatrix}$ & 
$m_\nu^3=m_0\begin{pmatrix}1 & y_1 & 0\\ y_1 & 0 & y_2\\0 & y_2 & y_3e^{i\alpha}\end{pmatrix}$ &
 $m_0=K_1,$ $K_2/K_1=y_1,$ $K_3/K_1=y_2,$ $K_4/K_1=y_3,$ $\alpha=(2k_2-2k_3+k_4-k_1)$\\
 \hline
 $m_\nu^4=\begin{pmatrix}K_1e^{ik_1} & 0 & K_2e^{ik_2}\\0 & K_4e^{ik_4} & K_3e^{ik_3}\\ K_2e^{ik_2} & K_3e^{ik_3} & 0\end{pmatrix}$ & 
$m_\nu^4=m_0\begin{pmatrix}1 & 0 & y_1\\ 0 & y_3 & y_2e^{i\alpha}\\y_1 & y_2e^{i\alpha} & 0\end{pmatrix}$ &
 $m_0=K_1,$ $K_2/K_1=y_1,$ $K_3/K_1=y_2,$ $K_4/K_1=y_3,$ $\alpha=\frac{1}{2}(k_1-2k_2+2k_3-k_4)$\\
 \hline
 $m_\nu^5=\begin{pmatrix}K_1e^{ik_1} & 0 & K_2e^{ik_2}\\0 & 0 & K_3e^{ik_3}\\ K_2e^{ik_2} & K_3e^{ik_3} & K_4e^{ik_4}\end{pmatrix}$ & 
 $m_\nu^5=m_0\begin{pmatrix}0 & 0 & y_1\\ 0 & 0 & y_2\\y_1 & y_2 & y_3e^{i\alpha}\end{pmatrix}$ &
 $m_0=K_1,$ $K_2/K_1=y_1,$ $K_3/K_1=y_2,$ $K_4/K_1=y_3,$ $\alpha=(k_1-2k_2+k_4)$\\
 \hline
 $m_\nu^6=\begin{pmatrix}K_1e^{ik_1} & K_2e^{ik_2} & 0\\K_2e^{ik_2} & K_4e^{ik_4} & K_3e^{ik_3}\\ 0 & K_3e^{ik_3} & 0\end{pmatrix}$ & 
$m_\nu^6=m_0\begin{pmatrix}1 & y_1 & 0\\ y_1 & y_3e^{i\alpha} & y_2\\0 & y_2 & 0\end{pmatrix}$ &
 $m_0=K_1,$ $K_2/K_1=y_1,$ $K_3/K_1=y_2,$ $K_4/K_1=y_3,$ $\alpha=(k_1-2k_2+k_4)$\\
 \hline
  $m_\nu^7=\begin{pmatrix} K_1e^{ik_1} & K_2e^{ik_2} & K_3e^{ik_3}\\K_2e^{ik_2} & 0 & K_4e^{ik_4} \\K_3e^{ik_3} & K_4e^{ik_4} & 0\end{pmatrix}$ & 
$m_\nu^7=m_0\begin{pmatrix}y_1 & 1 & y_2\\1 & 0 & y_3 e^{i\alpha}\\y_2 & y_3 e^{i\alpha} & 0\end{pmatrix}$ &
  $m_0=K_2,$ $K_1/K_2=y_1,$ $K_3/K_2=y_2,$ $K_4/K_2=y_3,$ $\alpha=(k_1-k_2-k_3+k_4).$\\
 \hline
\end{tabular}
\end{center}
\end{table}
We parametrize all emerged viable $m_\nu$ matrices in Table \ref{t20} in a generic way where $K_i$ and $k_i$ are functions of the elements of $m_D$, $m$ and $\mu_S$. We are not listing explicit expressions of each $K_i$ and $k_i$ parameters as there are many different functions for $K_i$ and $k_i$.

\subsection{Numerical analysis}

The matrix $m_\nu$ obtained in Linear seesaw case Eqn.\eqref{a1} is similar to the matrix $m_\nu^7$ obtained in Inverse seesaw case. In order to perform the numerical analysis we use the experimental constraints (Table \ref{data}) arising from the global fit oscillation data.
\begin{table}[!h]
\caption{Input experimental values\cite{Forero:2014bxa}}\label{data}
\begin{center}
\begin{tabular}{|c|c|}
\hline 
Quantity & 3$\sigma$ ranges \\ 
\hline 
$|\Delta m_{31}^2|$ (N)& 2.30$< \Delta m_{31}^2(10^3 eV^{-2})<2.64$ \\ 
\hline 
$|\Delta m_{31}^2|$ (I)& 2.20$< \Delta m_{31}^2(10^3 eV^{-2})<2.54$ \\ 
\hline 
 $\Delta m_{21}^2$& 7.11$< \Delta m_{21}^2(10^5 eV^{-2})<8.18$ \\ 
\hline 
$\theta_{12}$ & $31.8^o<\theta_{12}<37.8^o$ \\ 
\hline 
$\theta_{23}$ &  $39.4^o < \theta_{23}<53.1^o$  \\  
\hline 
$\theta_{13}$ &  $8^o < \theta_{13}< 9.4^o$  \\ 
\hline 
\end{tabular} 
\end{center} 
\end{table} We note that the first two matrices ($m_\nu^1$ and $m_\nu^2$) of Table \ref{t20} do not trigger $\beta\beta_{0\nu}$ decay, due to $|m_{11}|=0$ for these two matrices. Therefore, we categorize all the matrices presented in Table \ref{t20} into two different classes.\\
\noindent
\textbf{Class I:} Parameter ranges for allowed $m_{\nu}$ with $|m_{11}|\neq0$

For the numerical analysis of the matrices  $m_\nu^3,m_\nu^4,m_\nu^5,m_\nu^6$ we use the experimental constraints (Table \ref{data}) arising from the global fit oscillation data. It is seen that all the parameters are constrained in a very narrow range and we present them in Table \ref{t21}. The matrices predict a constrained range of $\delta_{CP}$ phase along with an upper bound on the sum of three light neutrino masses ($\Sigma_i m_i$) well below the upper limit dictated by the PLANCK and other astrophysical experiments\cite{Ade:2015xua}. For all the four matrices we get normal hierarchical spectrum of neutrino masses. The value of $m_{11}$ are also far below the present experimental probing region\cite{Auger:2012ar} .

\begin{table}[!h]
\caption{Parameter ranges of the matrices with $|m_{11}|\neq0$} \label{t21}
\begin{center}
\begin{tabular}{|p{0.5cm}|p{4.0cm}|p{2.5cm}|p{2.5cm}|p{2.5cm}|p{2.5cm}|}
\hline
$m_\nu$ & $y_1,y_2,y_3$ & $|\delta_{CP}|$ (deg) & $J_{CP}\times 10^3$ & $\sum m_i$  (eV) & $|m_{11}|$  (eV)$\times 10^2$\\
\hline
$m^3_\nu$ & \pbox{4.0cm} {$y_1:0.06-0.125$,\\$y_2:1.11-1.23$,\\$y_3:0.24-0.50$} & $3.96-5.25$ & $2.3-3.6$ & $0.146-0.215$ & $4.2-6.8$\\
\hline
$m^4_\nu$ & \pbox{4.0cm} {$y_1:0.06-0.23,$\\$y_2:1.118-1.386,$\\$y_3:0.259-0.866$} & $6.51-7.65$ & $3.8-4.8$ & $0.116-0.210$ & $3-6.4$\\
\hline
$m^5_\nu$ & \pbox{4.0cm} {$y_1: (7.98-8)\times 10^{-2},$\\ $y_2: 1.15-1.18,$\\ $y_3: 0.39-0.41$} & $9.0-9.4$ & $5.25-5.27$ & $0.14-0.172$ & $4.8-5.1$\\
\hline
$m^6_\nu$ & \pbox{4.0cm} {$y_1: 0.11-0.14,$\\ $y_2: 1.17-1.27,$\\ $y_3: 0.40-0.66$} & $5.72-7.53$ & $1.29-2.59$ & $0.128-0.173$ & $3.5-5.1$\\
\hline
$m^7_\nu$ & \pbox{4.0cm} {$y_1: 1.30-1.34,$\\ $y_2: 0.85-0.89,$\\ $y_3: 0.79-0.82$} & $0$ & $0$ & $0.127-0.131$ & $0.022-0.023$\\
\hline
\end{tabular}
\end{center}
\end{table}
\noindent
\newpage
\textbf{Class II:} Parameter ranges  of the matrices with $|m_{11}|=0$

\begin{figure}[h!]\begin{center}
\includegraphics[scale=0.4]{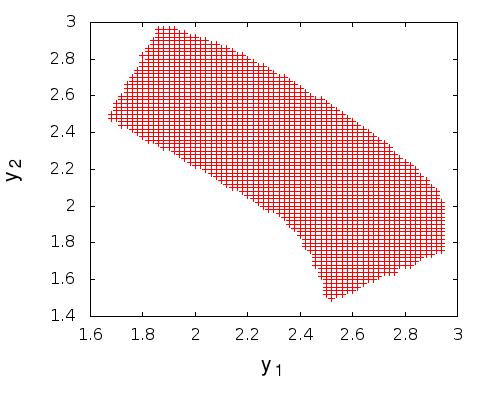}\includegraphics[scale=0.4]{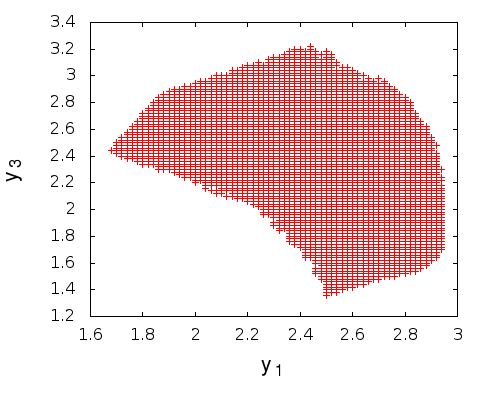}\\
\includegraphics[scale=0.4]{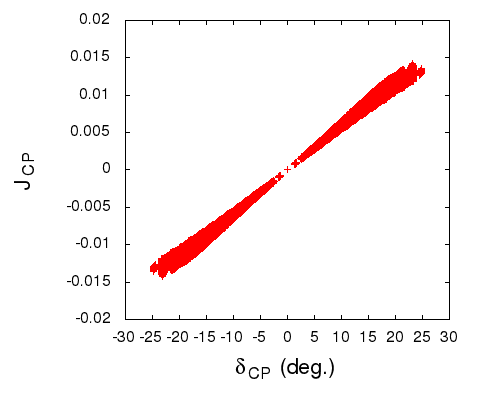}\includegraphics[scale=0.4]{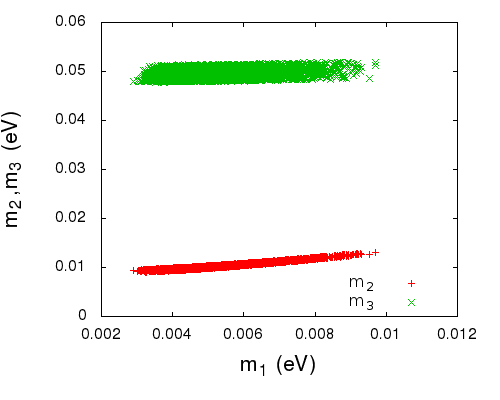}
\caption{The first two figures of the top row represent the parameter space for $m_\nu^1$ matrix. Left plot of the bottom row is the variation of $J_{CP}$ with $\delta_{CP}$ and the right figure shows the hierarchy (normal) of the model.}\label{f1}
\end{center}\end{figure}

\begin{figure}[h!]\begin{center}
\includegraphics[scale=0.4]{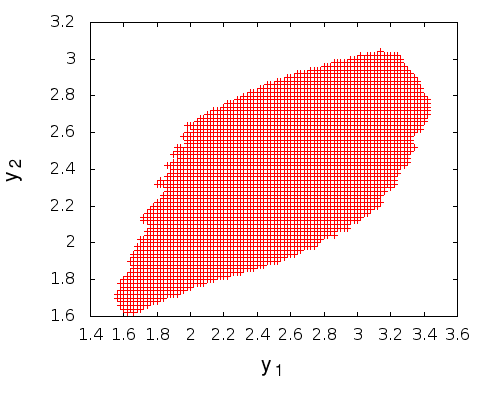}\includegraphics[scale=0.4]{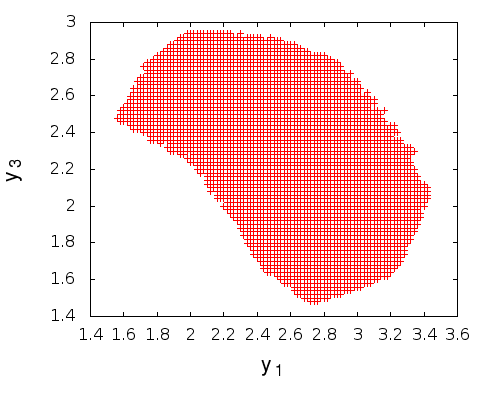}\\
\includegraphics[scale=0.4]{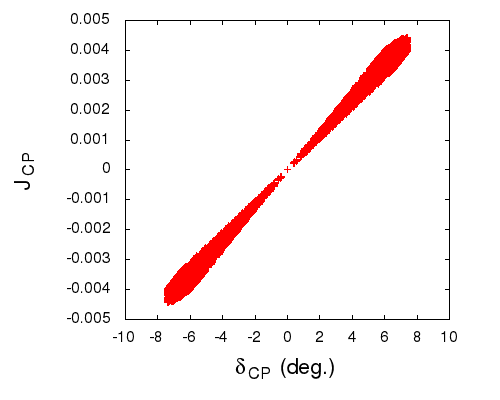}\includegraphics[scale=0.4]{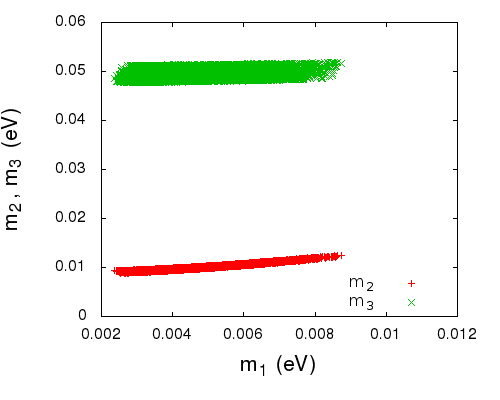}
\caption{The first two figures of the top row represent the parameter space for $m_\nu^2$ matrix. Left plot of the bottom row is the variation of $J_{CP}$ with $\delta_{CP}$ and the right figure shows the hierarchy (normal) of the model.}\label{f2}
\end{center}\end{figure}

Unlike the previous case, this class of matrices ($m_\nu^1$ and $m_\nu^2$) allow a sizable parameter space compatible with the experimental data. However, the matrices also predict constraint ranges of $\delta_{CP}$ phase and $\Sigma_im_i$. We present plots of these parameters in figure \ref{f1} and figure \ref{f2} respectively. From the first two plots of figure \ref{f1} the  ranges of the parameters read as $1.69<y_1<2.93$, $1.47<y_2<2.97$ and $1.37<y_3<3.16$. \\

The Dirac CP phase is constrained as $-25^o<\delta_{CP}<25^o$ and the sum of the light neutrino masses ($\Sigma_im_i$) is obtained within the range 0.094 eV $<\Sigma_im_i<$ 0.18 eV which is well below the present experimental upper bound. In figure \ref{f2} we present the parameter ranges for $m_\nu^2$. The matrix $m_\nu^2$ also allow a sizable parameter space and are depicted in first two plots of figure \ref{f2}. The ranges of $y_1$, $y_2$ and $y_3$ can be read as $1.58<y_1<3.4$, $1.5<y_2<3$ and $1.5<y_3<2.96$. Similar to the previous case, for this matrix also the ranges for $\delta_{CP}$ and $\Sigma_im_i$  are constrained in a very narrow range as $-8^o<\delta_{CP}<8^o$,  0.09 eV $<\Sigma_im_i<$ 0.16 eV. The hierarchy is normal and is depicted in the extreme right plot of the bottom row of figure \ref{f2}.\\



\section{Summary}\label{s5}

We analyze two low energy seesaw (Linear seesaw and Inverse seesaw) mechanisms with the assumption of a minimal non-singular structure of the charged lepton mass matrix $m_e$ with three distinct eigenvalues and non-zero eigenvalues for the effective neutrino mass matrix. Non-singular nature of $m_e$ and $m_\nu$ dictates certain possible textures for the constituent matrices. In the Linear seesaw, in our minimalistic approach,it is seen that 5 is the maximal number of zeros that can be accommodated in matrix `$m$' to obtain phenomenologically viable $m_\nu$. On the other hand, in the inverse seesaw, all the allowed two-zero textures can be explicitly realized in terms of the minimally parametrized constituent matrices. We have numerically explored the allowed parameter ranges using neutrino oscillation global fit data and predict $\sum m_i$, $|m_{11}|$, $J_{CP}$ and $\delta_{CP}$ along with the hierarchical structure of neutrino masses. One of the important prediction of this scheme is the vanishingly small value of $\delta_{CP}$ which could be tested by the ongoing T2K experiment. All the matrices predict nonvanishing  and highly constrained range of $\delta_{CP}$ along with the normal hierarchical spectrum of neutrino masses. 
Numerical analyses shows that two-zero textures cannot give rise to large CP violation, and therefore if $\delta_{CP}=\pi/2$ is established, this minimal scheme will be ruled out. However, we can possibly continue to have the same scheme in the neutrino sector but with other nontrivial charged lepton mass matrices such that $h_e=m_em_e^\dag$ is not diagonal to obtain large CP-violating phase.

\appendix

{}

\end{document}